\documentclass[aps,prl,reprint,groupedaddress,amsmath,amssymb]{revtex4-2}

\usepackage{graphicx}
\usepackage{bm}
\usepackage{hyperref}

\usepackage{siunitx}

\begin{document}
	
	\title{Efficient transport controlled by biharmonic frictional driving}
	\author{Martin Maza-Cuello}
	\author{Diego Maza}
	\affiliation{%
		Departamento de Física y Matemática Aplicada, 
		Facultad de Ciencias, 
		Universidad de Navarra,
		E-31080 Pamplona, Spain
	}%
	
	\date{\today}
	
	\begin{abstract}
		Dry friction has been proposed as a rectifying mechanism allowing mass transport over a vibrating surface, even when vibrations are horizontal and unbiased.
		It has been suggested that the drift velocity will always saturate when the energy of the input oscillation increases, leading to a vanishing efficiency that would hinder the applicability of this phenomenon.
		Contrary to this conjecture, in this work we experimentally demonstrate that, by carefully controlling the forcing oscillations, this system can maintain a finite transport efficiency for any input energy.
		A minimal friction model explains the observed dependencies of the drift velocity on the signal parameters in the case of biharmonic base oscillations, which can be extended to obtain efficiency estimates for any periodic excitation.
	\end{abstract}
	
	
	\maketitle
	
	\emph{Introduction.---}Dry friction is intimately associated with energy losses. 
	It has been calculated that 23$\%$ of the world’s total energy consumption is related to uncontrolled contact forces \cite{Tribo}. 
	Despite this, dissipative mechanisms are also essential in locomotion at broad spatial scales, 
	from robot design strategies \cite{Robot, ruggiero_nonprehensile_2018} to the dynamics of microswimmers \cite{Swimmers, elgeti_microswimmers_2016}. 
	The use of nonlinear couplings between dissipation and external forces for rectifying purposes has been extensively studied for nano- and microscopic systems exhibiting ratchet dynamics \cite{hanggi_artificial_2009}.
	These strategies have been successfully transferred to underdamped systems, e.g. controlled surface vibrations can lead to the uphill transport of microliter-sized drops \cite{noblin_ratchetlike_2009} or, when coupled with regions of different friction, can lead to precise positioning of solid pieces \cite{van_hoof_experimental_2014}.
	A smart design of the frictional actuation can thus provide a useful mechanism for transporting any mass.
	
	Ratchet-like transport has been predicted to emerge in a system composed of a mass supported by an asymmetrically vibrating surface \cite{buguin_motions_2006}.
	As oscillations gain strength, friction saturates and provokes the mass to slip.
	These slipping ``bursts'' may have different duration in each direction, 
	thus promoting directional macroscopic transport \cite{buguin_motions_2006}.
	This mechanism can originate from noisy signals \cite{gennes_brownian_2005, baule_path_2010, goohpattader_experimental_2009} as well as from adding transversal bias to a simple sinusoidal oscillation, for instance, by using a tilted surface \cite{benedetti_angular_2012} or by vibrating it diagonally \cite{digilov_dynamic_2010}.
	Purely horizontal, deterministic, and asymmetric oscillations can also lead to transport even if they have zero mean
	\cite{
		buguin_motions_2006,
		fleishman_directed_2007, 
		viswarupachari_vibration_2012, 
		nath_directed_2022,
		hui_vibrational_2024}.
	In this latter case, a perturbative argument led to the conclusion that the mass drift would become independent of the amplitude of the input oscillation, thus reaching a plateau \cite{buguin_motions_2006} that would limit its practical utility. 
	
	However, this work experimentally demonstrates that when two harmonic input signals are adequately composed, the drift velocity grows indefinitely as the input energy increases.
	A simple stiction model of friction together with a geometrical argument accurately describes how the drift depends on the signal shape. 
	This approach can be applied to obtain an estimation of the transport efficiency under any periodic oscillation, when the signal is strong enough to provoke the mass to continuously slip.
	
	\begin{figure}[b]
		\includegraphics[width=0.9\columnwidth]{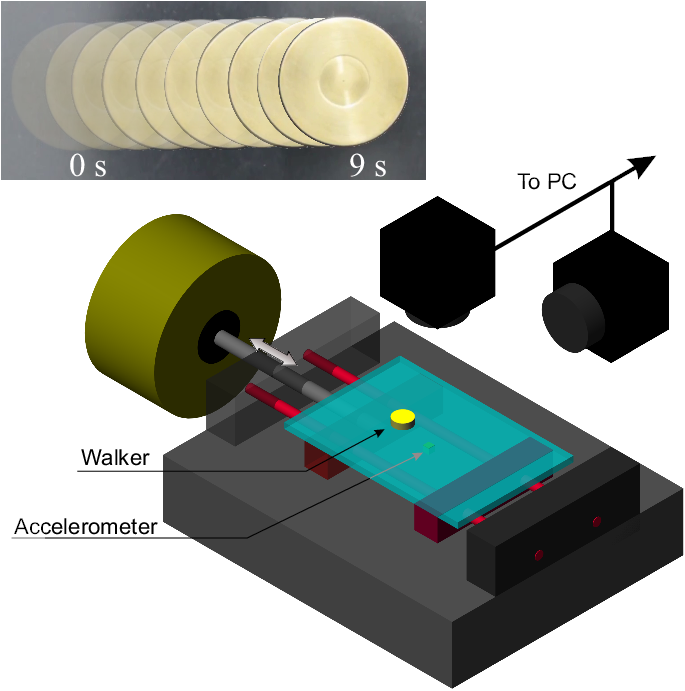}%
		\caption{\label{fig1}
			Sketch of the experimental setup. A free mass, or ``walker'', on top of a vibrating surface rectifies the oscillations and acquires a macroscopic drift velocity.
			A shaker connected to an arbitrary waveform generator provides horizontal oscillations.
			Two synchronized high-speed cameras record base and walker trajectories, and a 3-axis accelerometer synchronously registers the base acceleration. 
			Top: Composition of frames from Supplemental Video 1 \cite{SM}, showing the drifting motion of a brass walker (diameter: $\SI{3}{cm}$) under asymmetric oscillations of the base.
		}
	\end{figure}
	
	\emph{Macroscopic transport.---}The main components of our experimental setup are sketched in Fig.~\ref{fig1}.
	Technical specifications are provided in the Supplemental Material (SM) \cite{SM}.
	A free mass, or ``walker", lies on top of a base that oscillates horizontally. 
	The base is driven by a shaker connected to an arbitrary waveform generator to provide zero-mean oscillations. 
	Two synchronized high-speed cameras located on the side and top of the surface measure the base and mass positions.
	The input oscillation is simultaneously characterized by measuring the acceleration of the base $a_{\textrm{B}}(t)$ along the shaker axis via an accelerometer screwed to the base.
	The walker is a solid brass cylinder, indented on its bottom so that the contact area between mass and base is a polished thin ring less than $\SI{0.2}{\milli\meter}$ wide.
	Two different optical quality windows were explored as sliding surfaces: a fused quartz window and a flat optical diffuser (or frosted glass). 
	The same behavior was observed with each surface, results reported herein were obtained using the frosted glass.
	Walkers made of other materials, such as PMMA and aluminum, were explored and led to the same dynamics (see SM \cite{SM}), which is also largely independent on the walker's mass (see Supplemental Video 3 \cite{SM}).
	No specific surface treatment is necessary beyond thorough cleaning.
	
	In line with previously explored systems (both numerical 
	\cite{fleishman_directed_2007, 
		viswarupachari_vibration_2012, 
		nath_directed_2022} 
	and experimental \cite{hui_vibrational_2024}),
	we apply biharmonic oscillations as the driving signal, 
	which in general may be written as the family of base accelerations
	\begin{equation}
		a_{\textrm{B}}(t) = \gamma\left[\,\rho\sin(\omega t) + (1-\rho)\sin(2\omega t + \varphi)\,\right].
		\label{eq:a_b}
	\end{equation}
	Although $\gamma$ fixes the acceleration magnitude, the maximum acceleration is typically below this value, as $a_{\textrm{B}}$ depends on the (dimensionless) relative harmonic strength $\rho \in [0,1]$ and the phase $\varphi$.
	For the results in this Letter, the angular frequency is set at $\omega = \SI{125.66}{1\per\second}$ (oscillation period $T = \SI{50}{ms}$) and both harmonics have the same strength, $\rho = 1/2$. 
	Though the magnitude of the transport depends on $\rho$,
	very recent experiments with granular materials \cite{hui_vibrational_2024} indicate that $\rho = 1/2$ produces the maximum output drift for $\varphi = 0$, decaying smoothly for any other value. 
	Previous numerical studies have primarily focused on the $\varphi = \pi/2$ case 
	\cite{fleishman_directed_2007, 
		viswarupachari_vibration_2012, 
		nath_directed_2022}, 
	finding that the drift velocity saturates as the input amplitude is increased.
	This is in line with the original prediction about vibratory transport with asymmetric signals \cite{buguin_motions_2006}.
	Nevertheless, the effect that the shape of the input signal has on the resulting drift velocity has not yet been addressed experimentally.
	To rectify this, we study how the modulation of the biharmonic signal by controlling $\varphi$ affects the macroscopic drift.
	
\begin{figure}[t]
		\centering
		\includegraphics[width=\columnwidth]{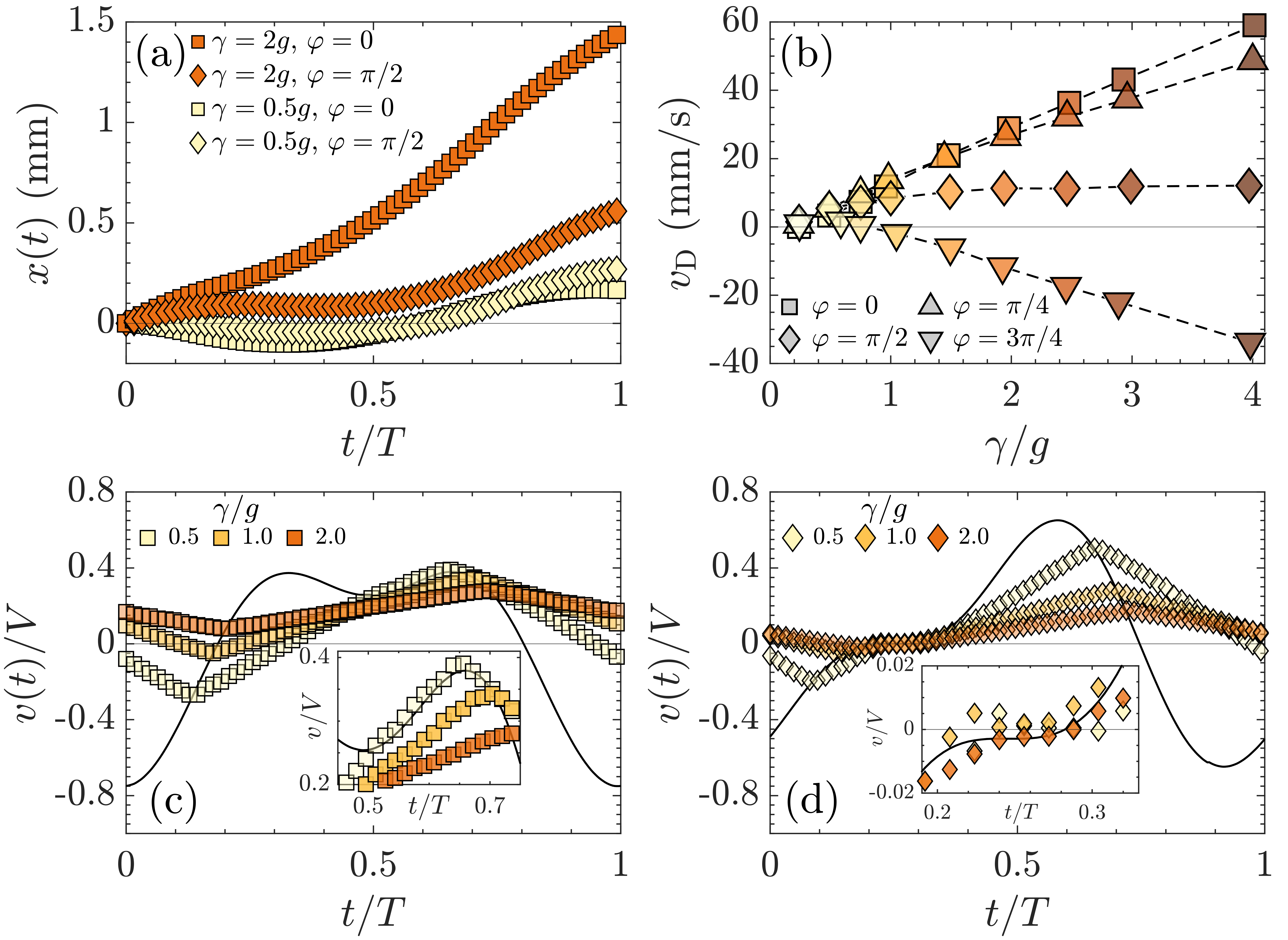}%
		\caption{\label{fig2}(a) Typical walker trajectories $x(t)$ over an oscillation period $T = \SI{50}{ms}$, obtained by averaging over 40 periods.
			See SM \cite{SM} for full trajectories.
			(b) Drift velocity $v_D$ as a function of $\gamma$ for indicated values of $\varphi$. Errorbars smaller than symbol size. 
			Dashed lines are guides to the eye. 
			(c) Walker velocities $v(t)$ during a period $T$ for $\varphi=0$, obtained from experimental trajectories such as the ones of panel (a), normalized by the velocity scale $V = \gamma/\omega$.
			(d) Same as panel (c) but for $\varphi=\pi/2$.
			In panels (c) and (d), insets are a zoom of ``stick'' intervals (see main text for details), and the black curve is the corresponding experimental base velocity $v_\textrm{B}(t)$ normalized by $V$.}
	\end{figure}

	The relevance of $\varphi$ is already patent when comparing averaged trajectories of the walker $x(t)$ during a single base oscillation period $T$ for different values of the phase, as exemplified in Figure~\ref{fig2}(a).
	Elapsed $T$, all trajectories acquire a certain drift $D = x(T) - x(0)$ due to the walker's slipping motion.
	The drift $D$ gained after each period $T$ is constant.
	Remarkably, as $\gamma$ increases from $0.5g$ to $2g$ ($g=\SI{9.8}{m\per\second\squared}$ being the gravitational acceleration) 
	the drift of the trajectories with $\varphi = \pi/2$ (diamonds) only doubles, whereas in the $\varphi = 0$ case (squares) the drift has nearly an 8-fold increase.
	Indeed, as shown in Fig.~\ref{fig2}(b), the drift velocity $v_\textrm{D} = D/T$ saturates for $\varphi = \pi/2$ (diamonds) but continuously increases in magnitude for all other phases.
	Surprisingly, some phases like $\varphi = 3\pi/4$ (inverted triangles), allow to invert the drift direction by \emph{increasing} the amplitude of the base oscillation (see Supplemental Video 2 \cite{SM}).
	
	The walker dynamics is better understood by looking at the velocity of the walker $v(t)$, computed from experimental trajectories.
	Figures~\ref{fig2}(c) and \ref{fig2}(d) show $v(t)$ normalized by the velocity scale $V = \gamma/\omega$, for fixed $\varphi = 0$ and $\varphi = \pi/2$ respectively.
	Indeed, the velocity always reaches a cycle with the same period as the base velocity $v_\textrm{B}(t)$ (black curve).
	These cycles can be qualitatively described as composed by the ``stick'' and ``slip'' intervals characteristic of dry friction models \cite{popov_contact_2010,pennestri_review_2016}.
	Near the drift threshold, in the $\varphi = 0$ case, Fig.~\ref{fig2}(c) ($\gamma = 0.5g$),
	$v$ diminishes until the relative velocity $v_{\textrm{R}} = v - v_{\textrm{B}}$ vanishes at around $t/T \simeq 0.15$.
	At this point the base acceleration is large and positive, prompting another slip step until $v$ becomes again equal to $v_\textrm{B}$ at around $t/T \simeq 0.5$.
	After that, the walker sticks to the base until $t/T \simeq 0.7$ (see inset), when it starts to gently slip in the opposite direction; the cycle resumes when $v_\textrm{R} = 0$ again. 
	As $\gamma$ increases, this stick interval shortens and disappears when $\gamma = g$. 
	However, when $\varphi = \pi/2$, Fig.~\ref{fig2}(d), a stick interval remains even for $\gamma = 2g$ ($0.24 \lesssim t/T \lesssim 0.28$, see inset), locking the following slip step to begin with a velocity close to zero.	
	Since $v_\textrm{D}$ can be identified with the average value of $v(t)$ during a period $T$, its saturation is connected with this locking.
	In contrast, for $\varphi = 0$ the unconstrained increase of the minimum of $v$ with $\gamma$ leads to $v(t)$ being always positive.
	
	The drifting dynamics discussed in Figure~\ref{fig2} are experimentally found for other surface/material combinations in the same parameter range \cite{SM}, 
	indicating that these are generic phenomena under the biharmonic signal and are not determined by the materials under consideration. 
	Of course, modeling the full dynamics would require knowledge on how the friction force depends on the relative velocity $v_{\textrm{R}}$. 
	This dependency is not, however, easily established when $|v_{\textrm{R}}| \rightarrow 0$ \cite{popov_contact_2010, pennestri_review_2016}.
	In the next paragraphs, we will approach the dynamics from a simplified geometrical perspective based on the cyclic nature of the walker velocity $v(t)$, 
	with the specific aim to isolate how the shape of the oscillations determine $v_\textrm{D}$.
	
	\emph{No-Stick approximation.---}
	Figure~\ref{fig2} shows that when $\gamma>g$, stick intervals become negligible and slip intervals are constrained by instants at which $v_\textrm{R}$ vanishes.
	Accordingly, we will describe the walker dynamics with an Amontons-Coulomb model \cite{pennestri_review_2016} depending on the relative velocity $v_\textrm{R}$ and characterized by a unique frictional coefficient $\mu$,
	\begin{equation}
		\dot{v}(t) = -\mu g\,\text{sgn}\left\{v_{\textrm{R}}(t)\right\}.
		\label{eq:NS_dyn}
	\end{equation}
	In writing Equation~\ref{eq:NS_dyn}, it is assumed that the acceleration when $v_{\textrm{R}} = 0$ is sufficiently high to provoke the walker to slip once again and that the change in sign is instantaneous. 
	Under these conditions, the walker never sticks, and its velocity cycle is piece-wise linear.
	The walker performs two opposing slip steps per period (see Figs.~\ref{fig2}(c) and \ref{fig2}(d)), 
	each of which is assumed to last half a period, $T/2$, when stick intervals are neglected. 
	Hence, Eq.~(\ref{eq:NS_dyn}) can be rewritten as a No-Stick (NS) approximation where the effective dynamics is given by a square acceleration
	\begin{equation}
		\dot{v}^{\textrm{NS}}(t) = \mu g\,\text{sgn}\left\{\sin(\omega[t-t_0])\right\}.
		\label{eq:vNS}
	\end{equation}
	The information about the driving oscillation is hidden in the shifting time $t_0$:
	at this instant, the velocity $v$ is required to be equal to the base velocity $v_{\textrm{B}}$, i.e. $v_{\textrm{R}}(t_0) = 0 = v_{\textrm{R}}(t_0 + T/2)$.
	Since the positive-slope step begins at $t_0$,  
	the shifting time is given by the minimum solution of
	\begin{equation}
		v_{\textrm{B}}(t_0 + T/2) - v_{\textrm{B}}(t_0) = \mu g \frac{T}{2}
		\label{eq:NS_t0}
	\end{equation}
	such that $\text{sgn}\left\{a_{\textrm{B}}(t_0 + T/2)\right\} = -\text{sgn}\left\{a_{\textrm{B}}(t_0)\right\}$ to guarantee opposite steps.
	Henceforth, $v_{\textrm{D}}$ is computed as the average between the starting velocities of each step,
	\begin{equation}
		v_{\textrm{D}}^{\textrm{NS}} = \frac{v_{\textrm{B}}(t_0) + v_{\textrm{B}}(t_0 + T/2)}{2} = v_{\textrm{B}}(t_0) + \frac{1}{4}\mu gT.
		\label{eq:NS_vD}
	\end{equation}
	Consequently, the drift velocity is a compromise between the initial velocity $v_\textrm{B}(t_0)$ and a constant drift arising from the duration of a slip step,
	both controlled by $\mu$ (recall that $t_0$ depends implicitly on $\mu$ through Eq.~(\ref{eq:NS_t0})).
	By construction, $v_\textrm{B}(t_0)$ determines the minimum of $v(t)$ during the NS cycle.
	Comparing with the experiment, Figs.~\ref{fig2}(c) and (d), 
	we see that Eq.~(\ref{eq:NS_vD}) links the growth of $v_\textrm{D}$ for $\varphi = 0$ to the continuous growth of $v(t)$, while the locking of $v(t)$ on the stick interval for $\varphi = \pi/2$ is the cause of the saturation of $v_\textrm{D}$.
	Interestingly, even if $v_{\textrm{B}}(t_0)$ saturates, an increase in $\mu$ will give a larger drift velocity.	
	
	In our experiments, the base velocity is
	\begin{equation}
		v_{\textrm{B}}(t) 
		= 
		-\frac{\gamma}{2\omega}
		\left[
		\cos(\omega t) + \dfrac{1}{2}\cos(2\omega t + \varphi)
		\right]
		\label{eq:vB}
	\end{equation}
	which, using Eqs.~(\ref{eq:NS_t0}) and (\ref{eq:NS_vD}), leads to the drift velocity 
	\begin{equation}
		v_{\textrm{D}}^{\textrm{NS}} = \frac{\gamma}{4\omega}\cos\big(\varphi - \Theta(\gamma)\big),
		\label{eq:vD}
	\end{equation}
	where the shift
	\begin{equation}
		\Theta(\gamma) = 2\text{asin}\left(\frac{\pi\mu g}{\gamma}\right)
		\label{eq:varphi_opt}
	\end{equation}
	determines the phase of maximum drift.
	
	\begin{figure}[h]
		\centering
		\includegraphics[width=\columnwidth]{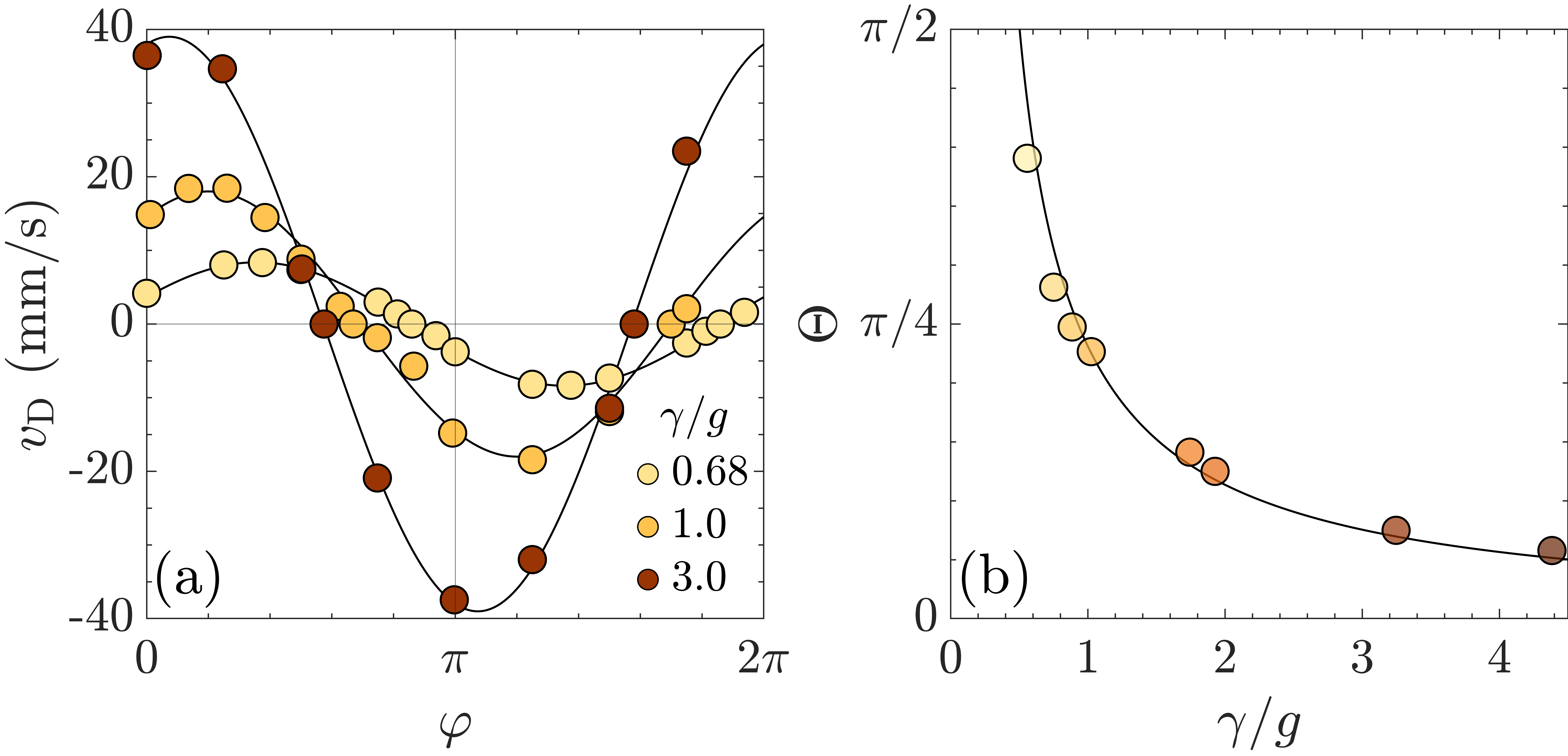}%
		\caption{\label{fig3} (a) $v_\textrm{D}$ vs. $\varphi$ for different $\gamma$ values.
			Curves are fits to $v_{\textrm{D}}(\varphi) = v_{\textrm{D}}^*\cos(\varphi - \Theta^*)$.
			For $\gamma = 0.68g$: $v_{\textrm{D}}^* = 8.3$ mm/s, $\Theta^* = 1.12$;
			for $\gamma = g$: $v_{\textrm{D}}^* = 17.6$ mm/s, $\Theta^* = 0.62$;
			for $\gamma = 3g$: $v_{\textrm{D}}^* = 38.7$ mm/s, $\Theta^* = 0.22$.
			(b) $\Theta$ as a function of $\gamma$.
			The curve is a fit to Eq.~(\ref{eq:varphi_opt}) giving $\mu = 0.113(6)$ (95\% confidence interval of the fit is given as error).
			In both panels, errorbars smaller than symbol size. 
		}
	\end{figure}
	
	The predicted harmonic dependence of $v_{\textrm{D}}$ on $\varphi$ is found to hold experimentally as shown in Fig.~\ref{fig3}(a), including the location of the maxima at a non-trivial phase $\Theta(\gamma)$ as expected from Eq.~(\ref{eq:vD}). 
	Measuring $\Theta$ is experimentally challenging,
	and instead we find the phase $\varphi_0(\gamma)$ such that $v_{\textrm{D}} = 0$, which can be determined with great precision.
	In Fig.~\ref{fig3}(b), we report $\Theta = \varphi_0 - \pi/2$, 
	since the harmonic dependency of Eq.~(\ref{eq:vD}) is verified experimentally.
	The decay of $\Theta(\gamma)$ is well described by Eq.~(\ref{eq:varphi_opt}), 
	allowing us to obtain an estimate of the dynamic friction coefficient $\mu \simeq 0.113$,
	coherently with other estimated values in our system \cite{maza-cuello}.
	$\Theta(\gamma)$ explains why $\gamma$ controls both the drift magnitude and its direction, as evidenced by the inverted triangles in Fig.~\ref{fig2}(b).
	
\begin{figure}[h]
		\centering
		\includegraphics[width=\columnwidth]{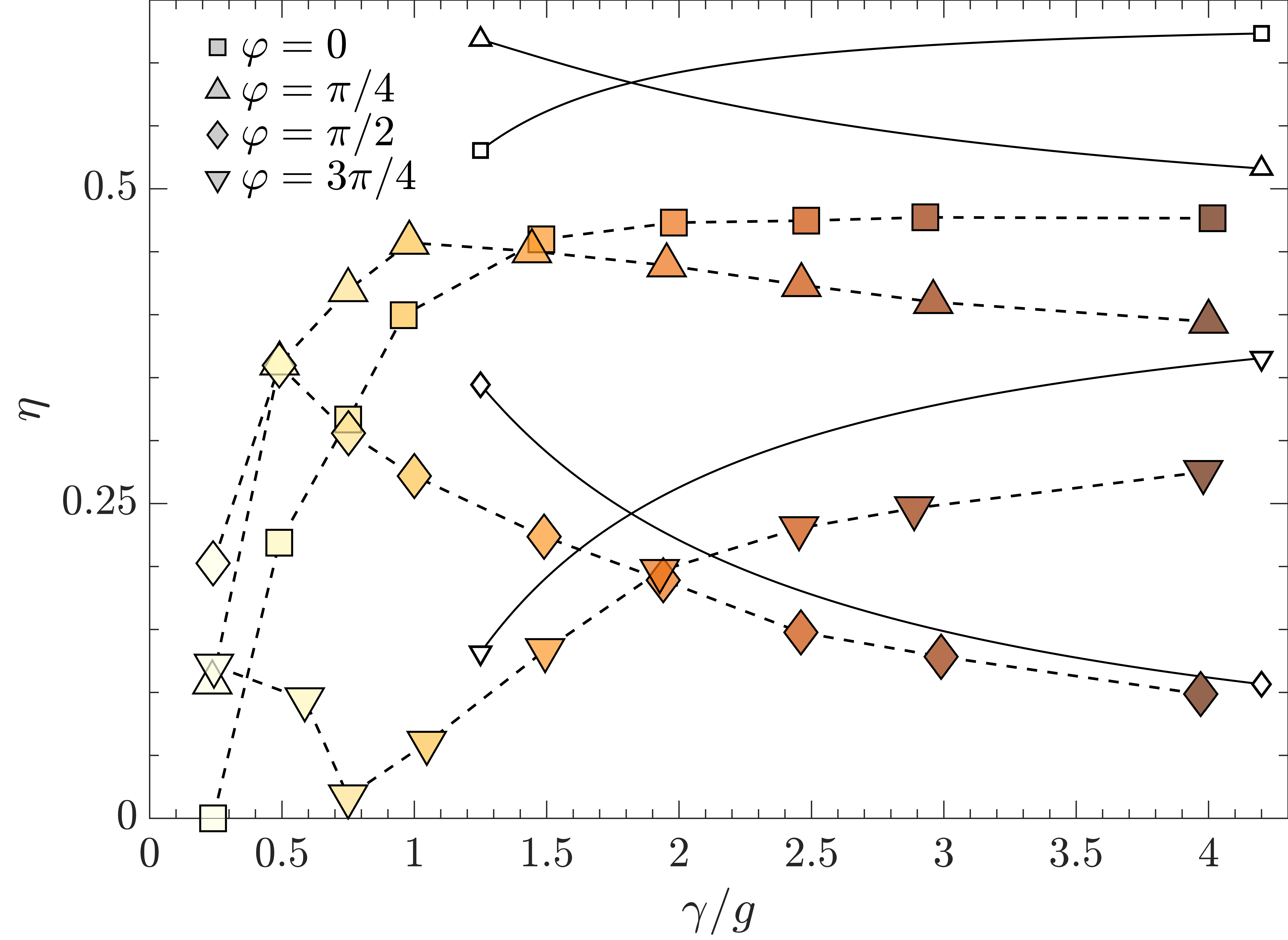}
		\caption{Filled symbols: transport efficiency $\eta = \left|v_\textrm{D}\right|/V_\textrm{RMS}$ corresponding to data reported in Fig.~\ref{fig2}(b), dashed lines are guides to the eye.
			Solid curves: predicted efficiency $\eta^{\textrm{NS}} = \left|v_\textrm{D}^{\textrm{NS}}\right|/V_\textrm{RMS}$ computed under the NS approximation via Eq.~(\ref{eq:vD}) with $\mu = 0.113$, empty symbols at endpoints label the phase.\label{fig4}}
	\end{figure}

	The magnitude of $v_\textrm{D}^\textrm{NS}$ predicted by Eq.~(\ref{eq:vD}) is about $30\%$ larger than the measured values of $v_\textrm{D}$ (cf. Fig.~\ref{fig4}). 
	This is caused not only by neglected stick intervals,
	but also by the finite relaxation time $\tau \sim \SI{2}{\milli\second}$ it takes to flip between slip steps (cf. Sup. Fig. 2 \cite{SM}),
	which is ruled by specific frictional details beyond the NS approximation.
	
	Despite these shortcomings, the geometrical construction leading to Eq.~(\ref{eq:vD}) is sufficient to describe the observed phenomena at medium and large $\gamma$ values.
	The same reasoning can be applied to more general periodic oscillations to obtain a first approximation of the drift velocity.
	For example, for the multiharmonic signal
	\begin{equation}
		v_{\textrm{B}}(t) = \dfrac{\gamma}{\omega}
		\left\{ \sum_{q = 1}^{n}a_q\cos(q\omega t + \alpha_q) \right\},
		\label{eq:v_bstar}
	\end{equation}
	with $a_q$ dimensionless weights and $\alpha_q$ phases, 
	the NS approximation predicts \cite{SM}
	\begin{equation}
		v_{\textrm{D}}^{\textrm{NS}} = \dfrac{\gamma}{\omega}\sum_{q \text{ even}}^{n}a_q\cos(q\omega t_0 + \alpha_q),
		\label{eq:vDgen}
	\end{equation}
	implying that a composition of odd harmonics cannot produce sustained drift, 
	similarly to microscopic ratchet systems with the same symmetries \cite{hanggi_artificial_2009}.
	
	\emph{Transport efficiency.---}Finally, we quantify the transport efficiency as $\eta = \left|v_\textrm{D}\right|/V_{\textrm{RMS}}$, where $V_{\textrm{RMS}}$ is the root mean square (RMS) velocity of the base, $V_{\textrm{RMS}} =\sqrt{\int_{0}^{ T}v^2_\textrm{B}dt/T} = \sqrt{5/32}\gamma/\omega$, where we have used Eq.~(\ref{eq:vB}). $V_{\textrm{RMS}}$ is independent of $\varphi$ and allows to compare signals with equal $\gamma$ but different geometries (the prefactor $\sqrt{5/32}$ arises from the different weights of the two components of $v_\textrm{B}$), and relates to the input energy.	
	Figure~\ref{fig4} shows $\eta$ computed from the data in Fig.~\ref{fig2}(b),
	demonstrating that non-vanishing efficiencies up to $\eta \approx 50\%$ can be obtained.
	The stark differences among the curves remark that the shape of the input oscillation is as essential as the input amplitude in obtaining the maximum possible transport.
	The NS efficiency $\eta^\textrm{NS} = \left|v_{\textrm{D}}^{\textrm{NS}}\right|/V$, although is higher than $\eta$ for the reasons explained above, 
	qualitatively follows the experimental trends, Fig.~\ref{fig4} (solid curves).
	Note that letting $\gamma \to \infty$ in Eq.~(\ref{eq:vD}) results in $\eta^\textrm{NS} \to \sqrt{2/5}\left|\cos\left(\varphi\right)\right|$, recovering a previously reported behavior in this limit \cite{hui_vibrational_2024}.
	This friction-independent behavior at large $\gamma$ values is expected as a consequence of the Amontons-Coulomb model given by Eq.~(\ref{eq:NS_dyn}) \cite{SM}.
	Therefore, $\varphi = \pi/2$ (and its counterpart $\varphi = 3\pi/2$) is the only phase resulting in a vanishing efficiency, 
	while all other phases are predicted to have finite $\eta$ 
	beyond the experimentally available $\gamma$ range.
	
	\emph{Conclusions.---}Dry friction can be used to rectify oscillations, allowing the transportation of a mass by vibrating the surface it stands on.
	In the case of a biharmonic signal, we experimentally show that it is possible to fine-tune the magnitude and direction of the induced macroscopic displacement by controlling the amplitude $\gamma$ and the phase difference $\varphi$ between the harmonics. 
	Although previously it has been assumed that the drift velocity would always saturate when increasing $\gamma$ \cite{buguin_motions_2006}, we experimentally find non-vanishing transport efficiencies of up to $\eta \approx 50\%$ when $\varphi$ is adequately chosen.
	
	Leaving aside the intrinsic complexity of surface interactions, we proved that an Amontons-Coulomb approach which only considers the geometry of slip steps can characterize the induced macroscopic displacement, quantified by the drift velocity $v_{\textrm{D}}$, when $\gamma$ is large enough for no significant stick intervals to exist.
	We derive an explicit expression for $v_{\textrm{D}}$ as a function of $\gamma$ and $\varphi$, which matches the experimental observations.
	$v_{\textrm{D}}$ grows with $\gamma$, but has a harmonic modulation with a phase that combines $\varphi$ and a $\gamma$-dependent shift $\Theta(\gamma)$ that also takes into account the dynamic friction coefficient $\mu$.
	An upper bound for $\eta$ can also be obtained,
	which qualitatively recovers the experimental trends in efficiency.
	Remarkably, the system behaves analogously to a microscopic inertial ratchet system such as the dissipation-driven current of the recoil of cold atoms in an optical lattice, cf. Figs.~3 and 4 in Ref.~\cite{gommers_dissipation-induced_2005}. 
	These results clarify the connection between microscopic ratchet dynamics \cite{hanggi_artificial_2009} and our macroscopic setup, 
	which has been hinted at since the first studies of friction as a rectifying force \cite{buguin_motions_2006}.
	
	Further understanding of vibratory transport will require detailed knowledge of frictional forces between the materials used.
	A critical factor in the optimization of the interplay between the applied signal and the resulting drift is the ability to control the stiction and pre-sliding process during the dynamics, 
	which may lead to increased efficiency for robotic manipulation and transport protocols \cite{ruggiero_nonprehensile_2018}.
	Moreover, considering other experimental setup parameters (e.g. chemical surface properties, adding lubricants, etc.) it should be possible to generate gains not only in transport efficiency, but also in more complex processes such as particle segregation or demixing.
	
	\begin{acknowledgments}
		We feel indebted to the Referees for their fruitful suggestions.
		We thank Niurka R. Quintero for inspiring discussions at the initial stages of the investigation.
		This work is supported by Grant PID2020-114839GB-I00 funded by MCIN/AEI/10.13039/501100011033.
	\end{acknowledgments}


\begin{thebibliography}{23}
\bibitem [{SM()}]{SM}%
	\BibitemOpen
	\href@noop {} {}\bibinfo {note} {See {S}upplemental {M}aterial at [URL will
		be inserted by publisher] for technical details of the experimental setup,
		{S}upplemental {F}igures and {V}ideos, and derivations within the {NS}
		approximation.}\BibitemShut {Stop}%
\end{thebibliography}
\end{document}